\documentclass{PoS}

\usepackage[small]{caption}
\usepackage{subcaption}
\usepackage{graphics,epstopdf,color,cancel,graphicx}
\usepackage[utf8]{inputenc}
\usepackage{lmodern}
\usepackage{textcomp}

\newcommand{\be}{\begin{equation}}
\newcommand{\ee}{\end{equation}}
\newcommand{\bary}{\begin{eqnarray}}
\newcommand{\eary}{\end{eqnarray}}
\newcommand{\en}{E_\nu}
\newcommand{\enc}{E_{\nu,c}}

\title{Observation of radio galaxies with HAWC}

\ShortTitle{HAWC Upper Limits for Radio Galaxies}

\author{\speaker{Daniel Avila Rojas}, Rubén Alfaro\\
        Instituto de Física, UNAM, Ciudad de México, México.\\
        E-mail: \email{daniel\_avila5@ciencias.unam.mx, ruben@fisica.unam.mx}}

\author{Antonio Galván, María Magdalena González, Nissim Fraija\\
        Instituto de Astronomía, UNAM, Ciudad de México, México.\\
        E-mail: \email{agalvan@astro.unam.mx, magda@astro.unam.mx, nifraija@astro.unam.mx}}

\author{Marc Klinger\\
        Department of Physics, Aachen University, Germany.\\
        E-mail: \email{marc.klinger@rwth-aachen.de}}

\author{for the HAWC Collaboration\\
        For a complete author list, see http://www.hawc-observatory.org/collaboration/\\
        }

\abstract{The High Altitude Water Cherenkov (HAWC) Gamma-Ray Observatory is an extensive air shower array located in Puebla, Mexico. The closest radio galaxy within the HAWC field of view, M87, has been detected in very high energies. In this work we report upper limits on the TeV $\gamma$-ray flux of the radio galaxy M87.  At a distance of 16 Mpc, M87 is a supergiant elliptical galaxy located in the Virgo Cluster that has been observed from radio wavelengths to TeV $\gamma$-rays. Although a single-zone synchrotron self-Compton model has been successfully used to explain the spectral energy distribution of this source up to a few GeV, the $\gamma$-ray spectrum at TeV has been interpreted within different theoretical models. We discuss the implications of these upper limits on the photo-hadronic interactions, as well as the number of neutrino events expected in the IceCube neutrino telescope.}

\FullConference{35th International Cosmic Ray Conference – ICRC217-\\
		10-20 July, 2017\\
		Bexco, Busan, Korea}

\begin{document}

\section{Introduction}

Active Galaxy nuclei (AGN) are one of the most powerful objects in the Universe. They are extraordinarily luminous across the entire electromagnetic spectrum, from radio to gamma rays. Radio galaxies, a type of AGN, are galaxies with non-thermal radio emission and lobes and jets emanating from the vicinity of the black hole. The spectral energy
distribution (SED) is usually described by leptonic and hadronic models. Leptonic models can explain emission up to GeV energy range by means of synchrotron self-Compton (SSC) emission \cite{2014MNRAS.441.1209F} and hadronic models at TeV energies by photo-hadronic processes \cite{2014A&A...562A..12P, 2014ApJ...783...44F}.  In accordance with the morphology, Fanaroff and Riley  proposed two classifications \cite{Fanaroff-Riley-74}. Class I with the bright radio emission close to its center and Class II with the radio emission peak further away. The radio galaxy M87 being one the four radio galaxies observed in TeV $\gamma$-rays is of interest due to its closeness to the Earth affording us an excellent opportunity for detecting very-high-energy (VHE) photons. It is worth noting that this radio galaxy is the closest one in the field of view of the High Altitude Water Cherenkov (HAWC) Observatory. HAWC is continually monitoring M87 and although no detection has been detected yet, upper limits on its VHE flux have been obtained.

In this paper, upper limits on VHE flux are reported and a hadronic model is presented in order to constrain the amount of protons in the jet and then, the number of high-energy neutrinos that could be detected by IceCube.  This work is arranged as follows. In Section 2, we introduce the HAWC detector. In Section 3, we give a brief description of the radio galaxy M87. In Section 4 and 5 we show the data analysis and upper limits to the VHE flux, respectively. In Section 6 the proposed hadronic model and the neutrino expectation are displayed and finally, brief conclusions are given in Section 7.

\section{The HAWC Observatory}

The High Altitude Water Cherenkov Observatory is an array of 300 Water Cherenkov Detectors (WCDs) located in Sierra Negra, Mexico at an altitude of 4100 meter above sea level. It is designed to detect extensive air showers (EAS) produced in the atmosphere by VHE gamma-rays and/or cosmic rays via the Cherenkov light generated by the secondary charged particles passing through the WCDs. Each WCD has a $7.3\,{\rm m}$ diameter tank with a $4.5\, {\rm m}$ depth, and is filled with $200,000$ liters of purified water.  In addition, each tank is instrumented with 4 Photomultiplier Tubes (PMT), a 10-inch PMT located at its center, and other three 8-inch PMTs located around it. HAWC is sensitive to gamma rays with energies in the range from $1\, $ to $100\, {\rm TeV}$ with a duty cycle $>95\%$ and an instantaneous field of view of $2\, {\rm sr}$. This observatory has already detected  gamma-ray fluxes from different sources \cite{2hwc-catalogue} such as the Crab Nebula \cite{hawc_crab}, Markarian 421 and Markarian 501 \cite{daily_monitoring} among others.

\section{M87}

M87 located near the center of the Virgo cluster at a distance of $16.7\, {\rm Mpc}$ ($z=0.0044$) \cite{2007ApJ...655..144M} is classified as Fanaroff-Riley Class I (FRI) source. The supermassive black hole in its nucleus has a mass of $M_{BH}\approx 3-6 \times 10^{9} M_{\odot}$ \cite{2013ApJ...770...86W}. It has a relativistic jet emerging from its nucleus that extends up to $2\,\, {\rm kpc}$ which has been studied in the X-rays by Chandra satellite \cite{2002ApJ...568..133W}. VHE emission has been detected by different detectors such as MAGIC, HESS and VERITAS \cite{2012A&A...544A..96A,2008ICRC....3..937B}. In particular, VERITAS reported a photon index of $2.31\pm 0.17_{stat}\pm 0.2_{sys}$ and a flux normalization of $7.4\pm  1.3_{stat}\pm 1.5_{sys}\times 10^{-13} {\rm cm^{-2}\,s^{-1}\,TeV^{-1}}$ \cite{2008ApJ...679..397A}.

\section{Data Analysis}

The likelihood method is used in order to estimate the significance of a source that has a low signal-to-noise ratio. The significance is directly related to the log-likelihood ratio by

\begin{equation}
TS\,=\,-2\ln\left(\frac{\mathfrak{L}_0}{\mathfrak{L}}\right)\,,
\label{tstatistic}
\end{equation}
where $TS$ is the test statistic, and $\mathfrak{L}_0$ and $\mathfrak{L}$ are the likelihood of the Null (no source model) and the Alternative (source model) hypotheses, which are modeled as a Poisson distribution of the event counts in each analysis and spatial bin $i$.  They can be written as

\begin{equation}
\mathfrak{L}_0  = \prod_i\ln\left(\frac{(B_i)^{N_i}e^{-(B_i)}}{N_i!}\right)\,,
\end{equation}
and
\begin{equation}
\mathfrak{L}  = \prod_i\ln\left(\frac{(B_i+S_i)^{N_i}e^{-(B_i+S_i)}}{N_i!}\right)\,, 
\end{equation}
where $S_i$ is the sum of the expected number of signal counts corresponding to a source with an specific spectra, $B_i$ is the number of background counts observed, and $N_i$ is the total number of counts observed. Therefore, $TS$ is given by

\begin{equation}
TS\,=\,\sum_i2\left[N_i\ln\left(1+\frac{S_i}{B_i}\right)-S_i\right]
\label{tstatistic_2}\,.
\end{equation}

To set a 95\% Confidence Level (CL) limit on the flux expected from the source we found the value of $S_i$ that maximizes the $TS$, $TS_{(max)}$, and then optimize $\Delta TS=TS_{(max)}-TS_{(95)}$, such that \cite{2017arXiv170601277A}

\begin{equation}
2.71 = TS_{(max)} - TS_{(95)} = TS_{(max)} - \sum_i2\left[N_i\ln\left(1+\frac{\xi S_i^{(ref)}}{B_i}\right)-\xi S_i^{(ref)}\right]
\label{cl_limit}\,.
\end{equation}

Here,  the number of expected signal counts from a source is scaled by a scale factor $\xi$, $S_i^{(ref)}$ is the expected number of signal counts in a bin calculated for a reference source spectral model and the flux normalization is $\langle F_{0} \rangle_{(ref)}$. The scale factor $\xi$ is then used to set a 95\% CL limit for a particular source
 
\begin{equation}
\langle F_{0} \rangle_{(95)}=\,\xi\times\langle F_{0} \rangle_{(ref)}
\label{sigma_95}\,.
\end{equation}

The limit is independent of the value chosen for $\langle F_{0} \rangle_{(ref)}$. The Likelihood method is implemented in the HAWC software utility \textsc{Liff} \cite{liff}.

\section{Upper Limits}

The relevance of the detection of VHE gamma-rays from this radio galaxy is that it could constrain the different models used to describe the particle accelerations in AGN as well as the region where this occurs. Within the AGN unification models it was thought that only blazars type were expected to be observed in the VHE emission band. The detection of variable VHE gamma-rays from M87 motivated the re-examination of the acceleration processes in non-aligned AGNs \cite{nonalign-AGN}.

Since the HAWC Observatory has not detected M87 with a statistical significance above $5\sigma$, upper limits for the flux normalization of a point like source were calculated for M87 using 760 days of data. It was also taken into consideration the effect of Extragalactic Background Light (EBL), so that the upper limit was calculated under two assumptions: one without EBL attenuation, since M87 is a nearby galaxy this is not an unlikely scenario\cite{Gil2012}; and the one which assumes the Franceschini EBL model \cite{2017arXiv170510256F}.

\begin{figure}[h!]
\captionsetup{width=0.8\linewidth}
\centering
\begin{subfigure}{1\linewidth}
\includegraphics[width=1.0\linewidth]{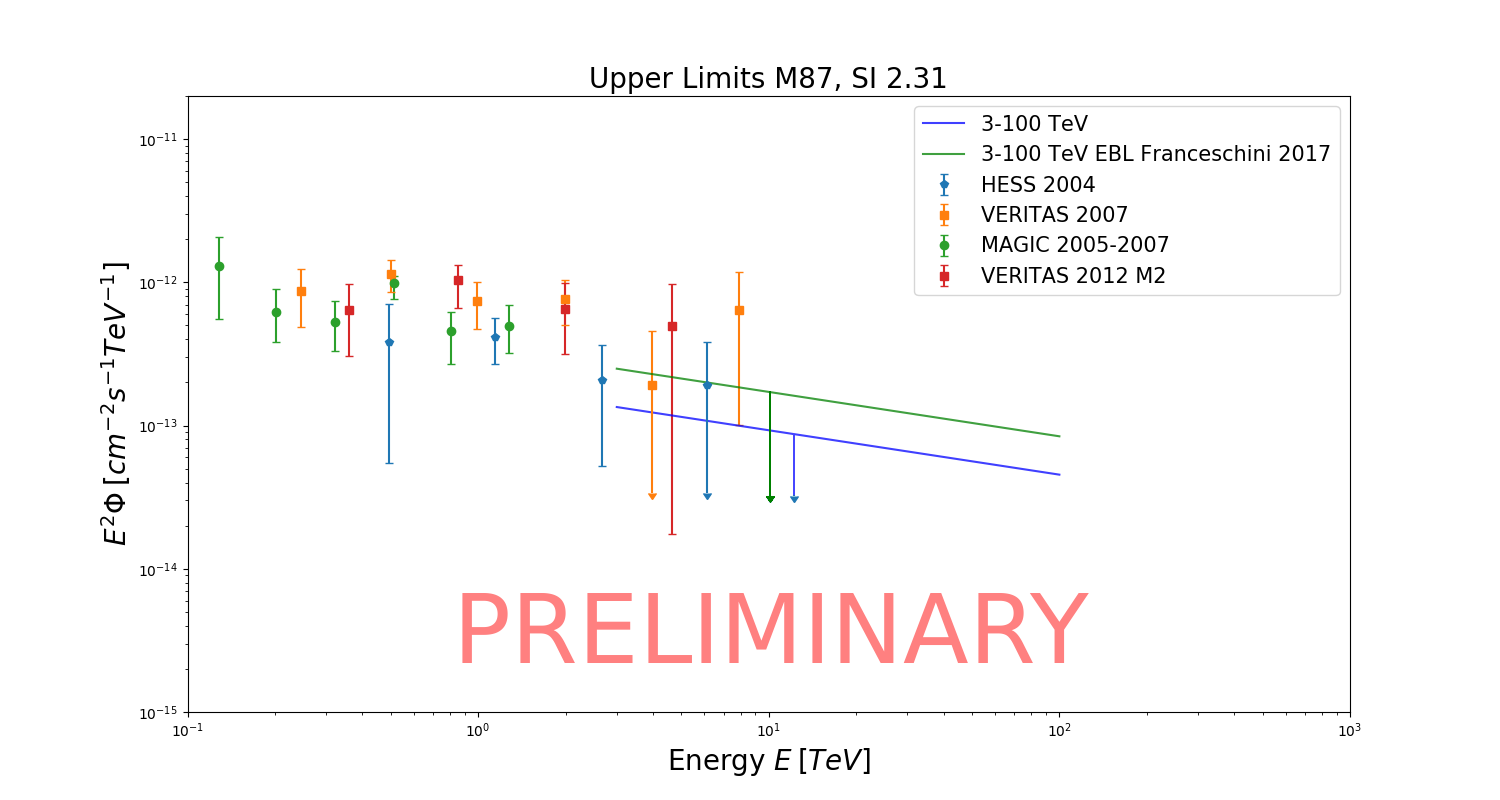}
\end{subfigure}
\caption{Upper limits calculated for M87. A comparison with the observations from MAGIC, HESS and VERITAS is shown.} 
\label{ul}
\end{figure}

Table \ref{ult} reports the flux normalization limit calculated with the likelihood method for M87. Figure \ref{ul} shows the comparison of the calculated upper limit with the observations of other experiments.

\begin{table}[h!]
\centering
\begin{tabular}{lcc}
\hline
Radio Galaxy & Upper Limit & UL with EBL\\
&[$10^{-13}\,{\rm TeV^{-1}\,cm^{-2}\,s^{-1}}$] & [$10^{-13}\,{\rm TeV^{-1}\,cm^{-2}\,s^{-1}}$]\\
\hline
M87 & 1.89 & 3.51\\ \hline
\end{tabular}\caption{Upper limits calculated for M87 with and without EBL.}\label{ult}
\end{table}

The HAWC upper limit for M87 without accounting for the EBL is below the extrapolation from the other experiments observations at energies beyond $1\;TeV$. This could mean that M87 is in a very low state never detected before, or the existence of a cutoff on its spectrum. Results based on including the Franceschini EBL model are consistent with the observations but are not the more likely scenario because of the closeness and the strong EBL attenuation predicted by the Franceschini's model while observations of other AGNs may print to a weaker EBL attenuation \cite{2013ApJ...770...77D}. The most recent observation by other experiments is the one made by VERITAS on 2012. All previous observations were made when M87 was in a low flux state. Because HAWC sees the average emission of the source, the upper limits account for possible flaring. 

\section{Hadronic Model}

Radio Galaxies have been proposed as a powerful accelerator of cosmic rays.  Accelerated protons can be described through  a simple power law \cite{2012ApJ...753...40F}

\be\label{prot_esp}
\left(\frac{dN}{dE}\right)_p=A_p\,E_p^{-\alpha_p}\,,
\ee
with $A_p$ the proportionality constant and $\alpha_p$ the spectral power index. The  proton density can be written as
\be
U_p=\frac{L_p}{4\pi\,\delta^2_D\,r^2_d}\,,
\ee
where $r_d$ is the emitting region, $\delta_D$ is the Doppler factor and $L_p$ is the proton luminosity which is given by 
\be\label{lum}
L_p= 4\,\pi\,d^2_z A_p\,\int \,E_p\,E_p^{-\alpha_p}  dE_p\,.
\ee
Accelerated protons loss their energies by electromagnetic channels and hadronic interactions. We consider that protons will be cooled down by p$\gamma$ interactions at the emitting region. Charged ($\pi^+$) and neutral ($\pi^0$) pions are obtained from p$\gamma$ interaction through the following channels   
\begin{eqnarray}
p \gamma &\longrightarrow&
\Delta^{+}\longrightarrow
\left\{
\begin{array}{lll}
p\,\pi^{0}\   &&   \mbox{fraction }2/3, \\
n\,  \pi^{+}      &&   \mbox{fraction }1/3\,,
\end{array}\right.
\end{eqnarray}
After that a neutral pion decays into photons, $\pi^0\rightarrow \gamma\gamma$,  carrying $20\% (\xi_{\pi^0}=0.2)$ of the proton's energy, $E_p$. The efficiency of the photo-pion production is \cite{1968PhRvL..21.1016S,1997PhRvL..78.2292W}
{\small
\begin{equation}\label{eficiency}
f_{\pi^0} \simeq \frac {t_{\rm dyn}} {t_{\pi^0}}  =\frac{r_d}{2\gamma^2_p}\int\,d\epsilon\,\sigma_\pi(\epsilon)\,\xi_{\pi^0}\,\epsilon\int dx\, x^{-2}\, \frac{dn_\gamma}{d\epsilon_\gamma} (\epsilon_\gamma=x)\,,
\end{equation}
}
where $t_{\rm dyn}$ and $t_{\pi^0}$ are the dynamical and the photo-pion timescales \cite{1997PhRvL..78.2292W}, respectively, $dn_\gamma/d\epsilon_\gamma$ is the spectrum of seed photons,  $\sigma_\pi(\epsilon_\gamma)$ is the cross section of pion production and $\gamma_p$ is the proton Lorentz factor. Taking into account that photons are released in the energy the range from $\epsilon_\gamma$ to $\epsilon_\gamma + d\epsilon_\gamma$ by protons in the energy range from $E_p$ to $E_p + dE_p$, then $f_{\pi^0}E_{p}(dN/dE)_{p}dE_{p}=\epsilon_{\pi^{0},\gamma}(dN/d\epsilon)_{\pi^{0},\gamma}d\epsilon_{\pi^{0},\gamma}$, then photo-pion spectrum is given by
{\small
\bary
\label{pgammam}
\left[\epsilon^2_\gamma \frac{dN_\gamma}{d\epsilon_\gamma}\right]_{\rm \gamma, \pi^0}= A_{\rm p\gamma}  \cases{
\left(\frac{\epsilon^{\pi^0}_{\gamma,c}}{\epsilon_{0}}\right)^{-1} \left(\frac{\epsilon_{\gamma}}{\epsilon_{0}}\right)^{-\alpha_p+3}          &  $ \epsilon_{\gamma} < \epsilon^{\pi^0}_{\gamma,c}$\cr
\left(\frac{\epsilon_{\gamma}}{\epsilon_{0}}\right)^{-\alpha_p+2}                                                                                        &   $\epsilon^{\pi^0}_{\gamma,c} < \epsilon_{\gamma}$\,,\cr
}
\eary
}
\noindent where the proportionality constant  $A_{\rm p\gamma}$  is in the form
\be\label{Apg}
A_{\rm p\gamma}= \frac{L_{\rm \gamma,IC}\,\sigma_{\rm \pi}\,\Delta\epsilon_{\rm res}\,\epsilon^2_0\,\left(\frac{2}{\xi_{\pi^0}}\right)^{1-\alpha_p}}{4\pi\,\delta_D^2\,r_d\,\epsilon_{\rm pk,ic}\,\epsilon_{\rm res}}\,A_p\,,
\ee
$\epsilon_0=$1 TeV is the energy normalization, $\epsilon_{\rm pk,ic}$ and $L_{\gamma,IC}$ are the energy and photon luminosity of the second SSC peak, repectively, $\Delta\epsilon_{\rm res}\approx$0.2 GeV, $\epsilon_{\rm res}\approx$0.3 GeV and  $\epsilon^{\pi^0}_{\gamma,c}$ is the break photon-pion energy given by {$\small \epsilon^{\pi^0}_{\gamma,c}\simeq 31.87\,{\rm GeV}\, \delta_D^2\, \left(\frac{\epsilon_{\rm pk,ic}}{ {\rm MeV}}\right)^{-1} $}.\\
%

Taking into consideration the upper limit derived in the previous section and the values of Doppler factor, photon Luminosity of the second peak, emitting region and the power index of proton distribution, derived  in \cite{2016ApJ...830...81F}, we found that the proton density and luminosity are less than 2.07 erg/cm$^3$ and $ 3.76 \times 10^{43}$ erg/s, respectively, when the EBL absortion is not considered and  3.83 erg/cm$^3$ and $ 6.96 \times 10^{43}$ erg/s when we consider the EBL absortion.

\subsection{High-energy neutrino expectation}
Photo hadronic interactions in the emitting region also generate neutrinos through the charged pion decay products ($\pi^{\pm}\rightarrow \mu^\pm+  \nu_{\mu}/\bar{\nu}_{\mu} \rightarrow  e^{\pm}+\nu_{\mu}/\bar{\nu}_{\mu}+\bar{\nu}_{\mu}/\nu_{\mu}+\nu_{e}/\bar{\nu}_{e}$).  Taking into account the distance of M87, the neutrino flux ratio (1 : 2 : 0 ) created on the source will arrive on the standard ratio (1 : 1 : 1 ) \cite{2014MNRAS.437.2187F, 2014ApJ...787..140F}. The neutrino spectrum produced by the photo hadronic interactions is
{\small
\bary
\label{pgammam}\label{espneu1}
\left[\en^2 \frac{dN_\nu}{d\en}\right]= A_\nu \epsilon^2_0 \cases{
\left(\frac{\en}{\epsilon_0 }\right)^2                            &  $ \en < \enc$\cr
\left(\frac{\en}{\epsilon_0 }\right)^{2-\alpha_{\nu}}     &   $\enc < \en$\,,\cr
}
\eary
}where the factor A$_{\nu}$ normalized through the TeV $\gamma$-ray flux is  {\small$A_\nu=A_{\rm p\gamma}\,\epsilon_0^{-2}\, 2^{-\alpha_p}$}.    
The number of neutrino events ($N_{\rm ev}$) expected in a detector considering a neutrino flux $\;dN_\nu/d\en$ during a period $T$, can be derived through the following relation \cite{2016ApJ...830...81F}
\bary\label{evtrate}
N_{\rm ev}&\simeq &T\,\int_{\en^{th}} A_{eff}(\en)\, \frac{dN_\nu}{d\en}\,d\en,
\eary
where $\en^{th}$ is the threshold energy and {\small $A_{eff}$} is the effective area\footnote{https://icecube.wisc.edu/science/data} of the instrument.\\

Taking into consideration that a neutral pion decays into two photons and a charged pion into three (anti)neutrinos and a lepton, and also that the three neutrino flavors ($\nu_e$, $\nu_\mu$ and $\nu_\tau$) are presented, then $\left[\en^2 \frac{dN_\nu}{d\en}\right]\simeq \left[\epsilon^2_\gamma \frac{dN_\gamma}{d\epsilon_\gamma}\right]_{\rm \gamma, \pi^0}$. From eq. (\ref{evtrate}) and the effective area of the IceCube telescope for a point-like source in the declination of M87 \cite{2014ApJ...796..109A}, we found that the upper limit on the number of neutrinos detected would be (considering electron, muon and tau neutrinos) $\sim 10^{-2}$. It is worth noting that the upper limit derived is far below the IceCube sensitivity flux \cite{2014ApJ...796..109A}. 

\section{Summary}

The $95\%$ CL upper limits on the flux normalization for the radio galaxy M87 were presented and discussed. The limits were calculated using 760 days of data collected with the HAWC Observatory. Within the proposed hadronic scenario we link the upper limits with the luminosity of accelerating protons in the jet. Considering the upper limits we constrain the amount of protons and then, the neutrino events expected in IceCube neutrino telescope. The HAWC observatory will continue monitoring M87 in order to  detect very-high-energy photons in the following years.

\section{Acknowlegments}

We acknowledge the support from: the US National Science Foundation (NSF); the US Department of Energy Office of High-Energy Physics; the Laboratory Directed Research and Development (LDRD) program of Los Alamos National Laboratory; Consejo Nacional de Ciencia y Tecnología (CONACyT), México (grants 271051, 232656, 260378, 179588, 239762, 254964, 271737, 258865, 243290, 132197), Laboratorio Nacional HAWC de rayos gamma; L'OREAL Fellowship for Women in Science 2014; Red HAWC, México; DGAPA-UNAM (grants RG100414, IN111315, IN111716-3, IA102715, 109916, IA102917); VIEP-BUAP; PIFI 2012, 2013, PROFOCIE 2014, 2015; the University of Wisconsin Alumni Research Foundation; the Institute of Geophysics, Planetary Physics, and Signatures at Los Alamos National Laboratory; Polish Science Centre grant DEC-2014/13/B/ST9/945; Coordinación de la Investigación Científica de la Universidad Michoacana. Thanks to Luciano Díaz and Eduardo Murrieta for technical support.

\end{document}